\documentclass{aa}
\usepackage{graphicx,supertabular}
\usepackage{txfonts}
\begin{document}
   \title{Luminous AGB stars in nearby galaxies}

   \subtitle{A study using Virtual Observatory tools}

   \author{P. Tsalmantza\inst{1}
          \and E. Kontizas\inst{2}
          \and L. Cambr\'esy\inst{3}
          \and F. Genova\inst{3}
          \and A. Dapergolas\inst{2}
          \and M. Kontizas\inst{1}}

    \offprints{P. Tsalmantza\\
    \email{vivitsal@phys.uoa.gr}}

    \institute{Department of Astrophysics Astronomy \& Mechanics, Faculty
               of Physics, University of Athens, GR-15783 Athens, Greece
         \and
              Institute for Astronomy and Astrophysics, National
              Observatory of Athens, P.O. Box 20048, GR-118 10 Athens, Greece
         \and
              Observatoire Astronomique de Strasbourg, F-67000 Strasbourg,
              France}

\date{Received date / accepted}

% \abstract{}{}{}{}{}
% 5 {} token are mandatory

  \abstract
   {} %leave it empty if necessary
   {This study focuses on very luminous ($M_{\rm bol}<-6.0$~mag) AGB stars with
   $J-K_s>1.5$~mag and $H-K_s>0.4$~mag in the LMC, SMC, M31, and M33 from 2MASS data.}
   {The data were taken from the 2MASS All-Sky Point Source catalogue archive.
   We used Virtual Observatory tools and took advantage of
   its capabilities at various stages in the analysis.}
   {It is well known that stars with the colors we selected correspond
   mainly to carbon stars. Although the most luminous AGBs detected here contain
   a large number of carbon stars, they are not included in existing
   catalogues produced from data in the optical domain, where they
   are not visible since they are dust--enshrouded. A comparison of the AGB
   stars detected with combined near and
   mid--infrared data from MSX and 2MASS in the LMC shows that 10\% of the
   bright AGB stars are bright carbon stars never detected before and that
   the other 50\% are OH/IR oxygen rich stars, whereas the 40\% that remain were
   not cross--matched.}
   {The catalogues of the most luminous AGB stars compiled
   here are an important complement to existing data. In the LMC, these
   bright AGB stars are centrally located, whereas they are concentrated in an active
   star--formation ring in M31. In the SMC and M33, there are not enough of them to draw
   definite conclusions, although they tend to be centrally located. Their
   luminosity functions are similar for the four galaxies we studied.
   {}

}

\maketitle

\section{Introduction}

The very red AGB stars, such as the various types of carbon and variable
stars in galaxies, trace the intermediate--mass stellar population, thereby providing
information about stellar evolution theories. Their spatial distribution is
related to the star--formation history of the parent galaxy. These reddest and
most extreme AGB stars cannot be detected easily at the conventional
$B$, $R$ wavelengths but can be observed in the near--infrared.

Carbon stars are detected in the optical domain if their mass is less than
4-5 $M_{\odot}$, whereas the more massive ones are expected to be seen only at
longer wavelengths. Dredge--up and mass--loss determine whether an AGB star
will become a carbon star. Very bright carbon stars can still be produced
from stars with masses of 6 or 8 $M_{\odot}$, if the mass--loss rate is not too
high to allow enough dredge--up episodes (Frost et al. \cite{frost}).
Models of stars up to 6 $M_{\odot}$  for solar and Magellanic metallicities
have shown that all massive carbon stars are, or could be,
dust--enshrouded and should therefore not be visible at $B$ and $R$
wavelengths (Frost et al. \cite{frost}). In the near--infrared domain,
where these stars are revealed best, carbon stars mainly populate the
{\em red tail} (Nikolaev \& Wenberg \cite{nikolaev} and Cioni et al.
\cite{cioni}), which is easily detected in the color--magnitude diagram
$M_{K_s}$ vs $J-K_s$ as an inclined branch of stars departing from an
almost vertical sequence of red giants. It is not yet clear whether
all stars in that area of the color--magnitude diagram are carbon stars.
Marigo et al. (\cite{marigo}) point out that the majority of {\em red tail}
stars are carbon, and Davidge (\cite{davidge2}) showed that, in NGC 205,
carbon stars are those among the {\em red tail} with the additional
constraints $J-K_s>1.5$~mag and $H-K_s>0.4$~mag.

Adopting the above criteria for carbon stars, we selected carbon
star candidates in the
Magellanic Clouds, M31 and M33 galaxies from the 2MASS All--Sky Survey.
The 2MASS detection limit allowed us to see only the brightest carbon star
candidates in M31 and M33. On the other hand, the brightest carbon dust--
enshrouded stars are not detected in the 4000-6000 \AA\ range in which the
published catalogues are based for both the SMC and LMC (Kontizas et al.
\cite{kontizas}; Rebeirot et al.  \cite{rebeirot}; Morgan et al.
\cite{morgan}). Our search for the most luminous AGB stars will produce a
homogeneous set of data for all four galaxies as confined by the detection
limit of 2MASS. Data retrieval and basic analysis were
performed through the Virtual Observatory (VO) and relevant tools.

Although the most luminous AGB stars are expected to be carbon stars, a lack
of carbon stars is observed for luminosities that are greater than
$M_{\rm bol}=-6.0$~mag (val Loon et al. \cite{vanLoon}).
One explanation is that luminous AGB stars become
invisible at wavelengths shorter than $\approx 1$~$\mu$m due to obscuration
by a circumstellar dust shell as a result of intense mass loss on the TP-AGB
(van Loon et al. \cite{vanLoon}).

In the LMC cluster HS 327, there is evidence that carbon stars and OH/IR stars may be
coeval (van Loon et al. \cite{vanLoon}). These stars are most likely
$\approx 200$~Myr old and formed at an epoch of intense star formation in the LMC.

Egan et al. (\cite{egan}), when combining 2MASS and MSX colors, studied
the AGB population and other red stars in the LMC, including objects with
unusual IR excesses. By using the IR point--source model by Wainscoat et al.
(\cite{wainscoat}) and obtaining source names and spectral types from the SIMBAD
database when available, they have identified 11 categories of stellar populations
and red nebulae, including main--sequence stars, giant stars, red supergiants (RSGs), C-- and
O--rich AGB stars, PNs, HII regions, and other dusty objects likely associated
with early--type stars. A total of 731 of these sources were previously unidentified.
Comparing their results with the bright AGBs of the LMC studied here, we found
that the majority of these stars are OH/IR and carbon stars.

In this study we provide the catalogues of the brightest AGB stars
($-8.4 < M_{\rm bol} < -6.0$~mag) in four galaxies: the LMC, the SMC, M31, and M33.
This selection of the magnitude range is due to the limit of detection of 2MASS in
M31 and M33. The spatial distribution and corresponding luminosity functions are
derived and discussed for each galaxy.

\section{Data Analysis}
\subsection{VO tools}
The recent developments of the Virtual Observatory creat
tools dedicated to some generic operations. Padovani et al.
(\cite{padovani}) prove the efficiency of such tools in helping astronomers
to produce scientific results using the European Astrophysical Virtual
Observatory (AVO).
We took advantage of the VO capabilities at various stages of the analysis
presented in this paper. For instance, the CDS service Vizier (Ochsenbein
et
al. \cite{ochsenbein}) is used as an implicit cross--match tool to identify
the LMC carbon stars from Kontizas et al. (\cite{kontizas}) in the 2MASS
point source catalogue. It is a positional cross--match that looks for
sources within a maximum distance chosen by the user (1\arcsec\ for this
work).
One of the main features of the VO is {\em interoperability}.
Vizier output for 2MASS queries is directly displayed using VOPlot
(Kale et al. \cite{kale}), a tool provided by the Indian VO. The
interoperability between Vizier and VOPlot is ensured by the
exchange of a VOTable\footnote{http://www.ivoa.net} that makes it possible
to directly visualize the spatial distribution or color--magnitude
diagrams resulting from the catalogue query. Moreover, the filtering
capabilities of VOPlot allow one to use criteria to select and display a specific
population, such as the AGB stars discussed in this paper.
As an example, Fig. \ref{f8} shows the superpositon of a color--filtered catalogue on
the M31 GALEX image (Thilker et al. \cite{thilker}). No WCS (World Coordinate System) information was initially
available, but a registration could be obtained using Aladin (Bonnarel et al. \cite{bonnarel}),
which was the basis of the AVO prototype. There are still limitations in the VO tools.
At the moment, for instance, very large catalogues cannot be handled in the same way as
small catalogues. While the M31 catalogue ($10^4$ sources) is directly output
in VOPlot, the LMC catalogue (2 million sources) is obtained via
{\em vizquery}\footnote{http://cdsweb.u-strasbg.fr/doc/vizquery.htx}, which
allows one to query VizieR remotely for further analysis.

\subsection{AGB stars and Bolometric correction}
The data used in this investigation were selected from the 2MASS all-sky
survey catalogue restricted to objects for which the error is lower than
0.15 mag in all bands.

There are several criteria for detecting carbon stars in the various wavelengths.
For {\ bf $JHK_s$} filters the most conservative criteria are those by Davidge
(\cite{davidge2}), who found that stars with
$H-K_s>0.4$~mag and $J-K_s > 1.5$~mag in the galaxy NGC 205 are carbon stars. These color limits
are known to be metallicity--dependent, with $J-K_s$ becoming redder by 0.1 mag
when the metallicity changes from Z=0.004 to Z=0.2 (Davidge \cite{davidge2}).
The metallicities of the four galaxies studied  here are either similar
or lower than the one studied by Davidge, which implies that the color $J-K_s$
can be bluer than the 1.5~mag value adopted here. If taking the metallicity effect into consideration,
our sample might not include some of the bluest stars.
Considering that the upper part of the red tail contains C-rich, O-rich, OH/IR stars,
only spectra would be able to solve the ambiguity about the nature of these bright red stars.

To compare the results found here with previous studies of the carbon star
populations in galaxies, the $K_s$ magnitudes of the stars were converted
into bolometric magnitudes. The bolometric correction used is an approximation
proposed by Bessel \& Wood (\cite{bessel2}). Some more recent references are
actually based on a similar assumption. Bessel \& Wood (\cite{bessel2})
computed the bolometric correction, $BC_{K}$, in the $K$ magnitude for
various AGB stars in the Galaxy, LMC, SMC, and 47 Tuc. The available carbon
star observations were too few to deduce a relation for them alone, but it was
found that the O-rich stars' relation for $M_{\rm bol}$ describes well the
few carbon stars in the $H-K$ vs $J-K$ diagram. In this same study it was shown
that the bolometric corrections vary for stars in galaxies with different
metallicities. For stars in the Galaxy and LMC, the correction given
by Bessel \& Wood (\cite{bessel2}) is $BC_{K}=0.72+2.65 \times (J-K)- 0.67 \times (J-K)^{2}$,
which was validated for the magnitude range $0.6<J-K<2.0$~mag. This is the
correction we used for the LMC, but also for M31 and M33, considering that they
have the same metallicity as the Galaxy and LMC respectively. For stars
in the SMC, the correction found by Bessel \& Wood (\cite{bessel2}) was
$BC_{K}=0.60+2.65 \times (J-K)-0.67 \times (J-K)^{2}$ for stars with
$0.6<J-K<1.5$~mag. Beyond that color limit no data were available in the SMC,
so in this case we used the bolometric corrections for stars in the LMC as an
approximation. This might lead to calculated values of $M_{\rm bol}$ that are higher than the real
ones for the SMC stars, but the difference between the two formulae is only
0.1 mag. This result is also supported by Montegriffo et al.
(\cite{montegriffo}) who studied the bolometric corrections in several
Galactic globular clusters with various metallicities. They found that
the bolometric corrections for metal-poor objects are very close to
those calculated from the metal-rich relation at redder colors.

\section{Color--magnitude diagrams}
\subsection{The Large Magellanic Cloud}
\begin{figure}
\centering
\includegraphics[width=9cm]{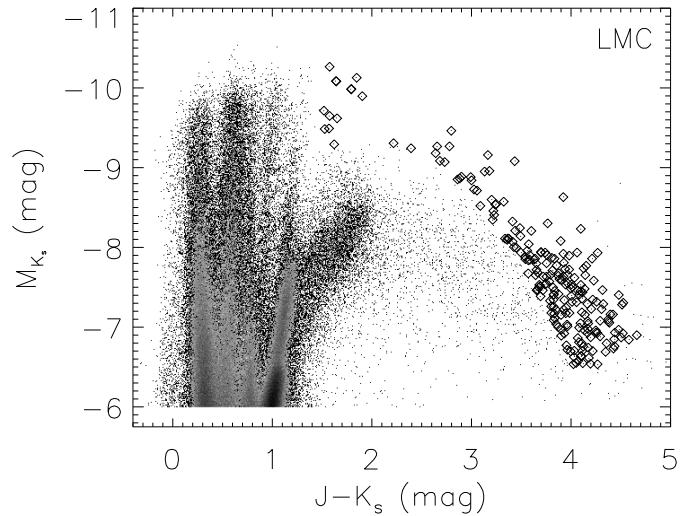}
\caption{Color--magnitude diagram for the LMC stars with $M_{K_s}<-6.0$~mag.
The {\em red tail} is defined by $J-K_s > 1.3$~mag and $M_{K_s} < -6.5$~mag,
whereas the luminous AGB stars with $-8.4 < M_{\rm bol} < -6.0$~mag are shown
with diamonds}.
\label{f1}
\end{figure}

Modeling the color--magnitude diagram of the {\em red tail} of the LMC from
the 2MASS data, Marigo et al. (\cite{marigo}) found that field stars are not
expected to reach colors that are redder than $J-K_s \approx 1.0$~mag, thus they
hardly contaminate the features produced by the LMC population of AGB stars.
We assumed the distance modulus $m-M=18.55 \pm 0.04 {\rm (statistical)} \pm
0.08 {\rm (systematic)}$~mag (Cioni et al. \cite{cioni}) and the mean reddening
$E_{B-V}=0.13$~mag (Massey et al. \cite{massey}), which gives an absorption of
$A_{K_s}=0.04$~mag using $A_{K_S}=0.34 \times E_{B-V}$
(Bessel \& Brett \cite{bessel}).
Carbon star candidates were selected following Davidge criteria (\cite{davidge2})
in the 2MASS $M_{K_s}$ vs $J-K_s$ diagram presented in Fig \ref{f1}.
Keeping only objects with 2MASS photometric errors less than 0.15~mag,
we found 7137 carbon star candidates among the 10055 stars
that belong to the {\em red tail}, which is defined by $J-K_s > 1.3$~mag
and $M_{K_s} < -6.5$~mag (Cioni et al. \cite{cioni}). The cross--match
within a 3\arcsec\ radius of the 7137 AGB stars with the 7716
spectroscopically confirmed carbon stars from Kontizas et al.
(\cite{kontizas}) gives 3782 common objects. This confirms that
{\em red tail} stars contain a large number of carbon stars.

\begin{figure}
\centering
\includegraphics[width=9cm]{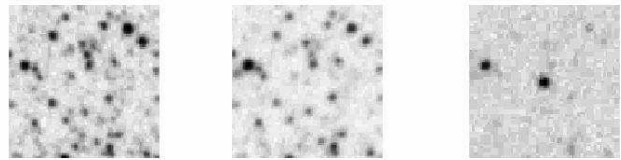}
\caption{The $B(J)$, $R$, $I$ images from the DSS survey of one LMC carbon
star detected here (2MASS J04532183-7051449) show that it is only
detected in the $I$ band.}
\label{f2}
\end{figure}

A comparison with Kontizas et al. (\cite{kontizas}) data shows that the
bright end of the optically detected carbon stars is two magnitudes fainter
in the near--infrared than the bright end of the {\em red tail} stars
detected in 2MASS. The bright AGB stars with
$-8.4 < M_{\rm bol} < -6.0$~mag are found to be 216 (Table \ref{lmc}), whereas
the total number of the bright {\em red tail} stars is 256. In addition, we
searched the DSS catalogue in three colors $B(J)$, $R$, and $I$ to confirm that
these stars are not observable in the $B$, $R$ wavelengths. We found that
many stars (almost 1 out of 5) are not visible in $B$ and $R$
(traditionally accepted as the representative optical window), but they
become detectable or appear very luminous in $I$, as shown in the
example of Fig. \ref{f2}. Deep surveys may reveal these stars, but
available observations have not detected them yet.

\begin{figure}
\centering
\includegraphics[width=9cm]{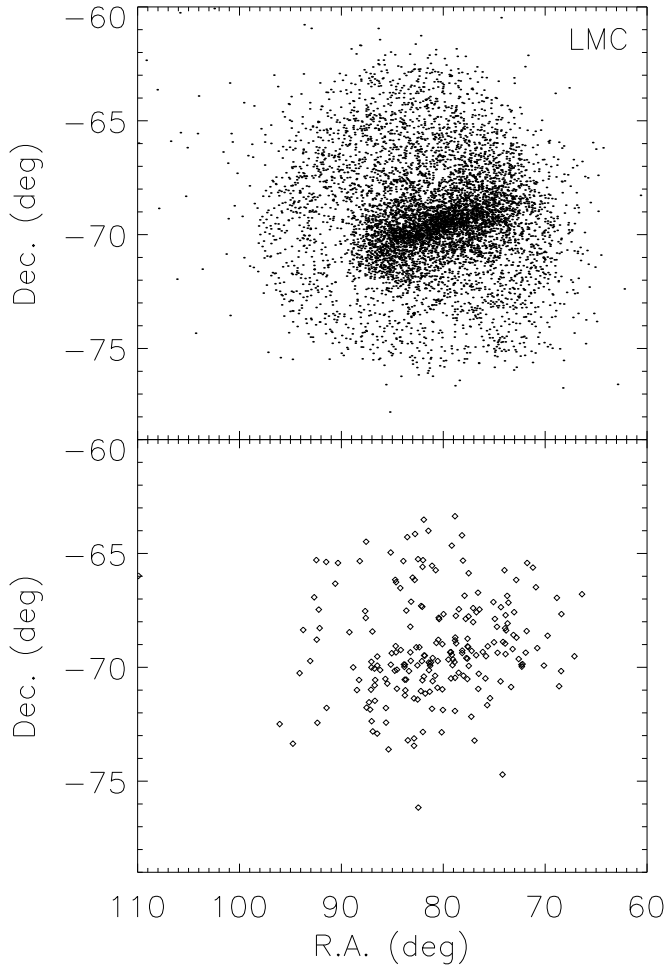}
\caption{Spatial distribution of all LMC AGB stars selected in this paper (top),
and the very bright ones with $-8.4 < M_{\rm bol} < -6.0$~mag (bottom).}
\label{f3}
\end{figure}

The spatial distribution of the 7137 {\em red tail} AGB stars extracted from
2MASS is presented at the top of Fig. \ref{f3}, and the bottom figure shows the
distribution of the 216 most luminous of them. Figure \ref{f3} shows high
density in the central regions for the
{\em red tail} AGB stars, including the most luminous ones, in agreement
with previous studies by Hughes \& Wood (\cite{hughes}) and Wood et al.
(\cite{wood}).

Egan et al. (\cite{egan}) combined LMC data from 2MASS and MSX to detect
carbon stars. They defined carbon
stars by using the criteria $1.5 < K_s-A < 3.75$~mag and
$0.8 < H-K_s < 1.5$~mag and the OH/IR with $J-K_s > 3$~mag,
$K_{s}-A > 3.75$~mag and $H-K_s > 1.5$~mag. Cross--matching the stars found here
with those of Egan et al. (\cite{egan}) has shown that 20 of the luminous AGB
stars are actually carbon stars, 109 of them are OH/IR stars, 5 are planetary
nebulae, 1 is oxygen--rich AGB and 1 is carbon--rich AGB star, 24 are of unkown nature,
and no match was found for the 56 other stars included in our sample.

\subsection{The Small Magellanic Cloud}
\begin{figure}
\centering
\includegraphics[width=9cm]{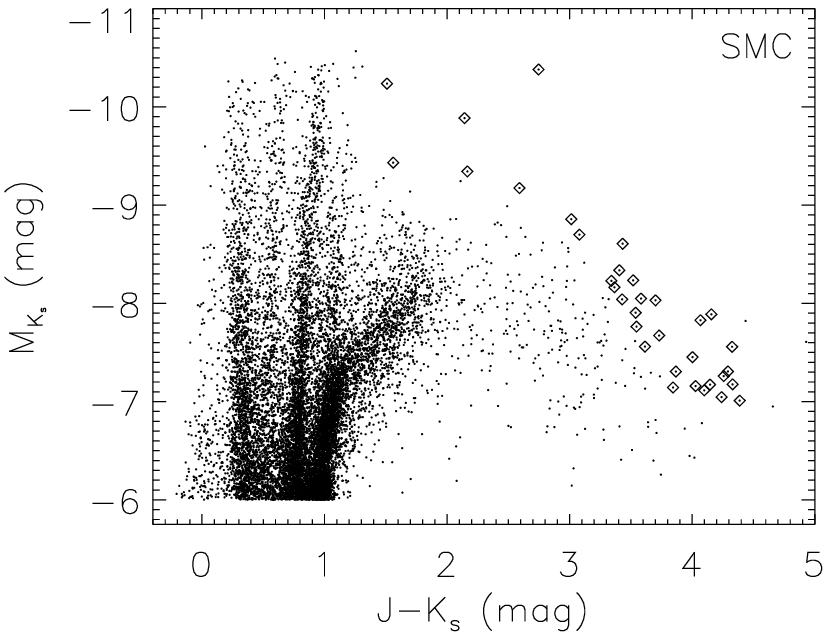}
\caption{Color--magnitude diagram for all SMC stars with $M_{K_s}<-6.0$~mag.
The {\em red tail} is defined by $J-K_s > 1.2$~mag and $M_{K_s} < -7.0$~mag,
whereas the luminous ABG stars with $-8.4 < M_{\rm bol} < -6.0$~mag are shown with
diamonds.}
\label{f4}
\end{figure}

Like the LMC, no significant foreground star contamination is expected at
$J-K_s>1.0$~mag (Marigo et al. \cite{marigo}) toward the SMC. The distance
modulus is $m-M=18.99 \pm 0.03 {\rm (statistical)} \pm 0.08 {\rm
(systematic)}$~mag (Cioni et al. \cite{cioni}). The mean reddening adopted
here is the one given by Massey et al. (\cite{massey}) $E_{B-V}=0.09$~mag,
which corresponds to $A_{K_s}=0.03$~mag.
The {\em red tail} known to be located at $J-K_s>1.2$~mag and $M_{K_s}<-7.0$~mag
can be seen in Fig. \ref{f4}. It contains 1674 stars in the 2MASS
catalogue when restricted to sources with photometric errors smaller than 0.15~mag.
Among them, 911 are carbon star candidates if assuming the Davidge (\cite{davidge2})
criteria.
There are 34 bright carbon stars with $-8.4 < M_{\rm bol} < -6.0$~mag (Table
\ref{smc}) out of a total of 50 bright sources in the {\em red tail}.
A 3\arcsec\ radius cross--match of the 911 carbon star candidates with the
catalogues of SMC carbon stars confirmed by spectroscopy in the optical
wavelengths (Rebeirot et al. \cite{rebeirot}; Morgan et al. \cite{morgan})
gives 676 sources in common. It is worth noting that only one cross--matched
star belongs to the 34 very bright AGB stars of our sample.

\begin{figure}
\centering
\includegraphics[width=9cm]{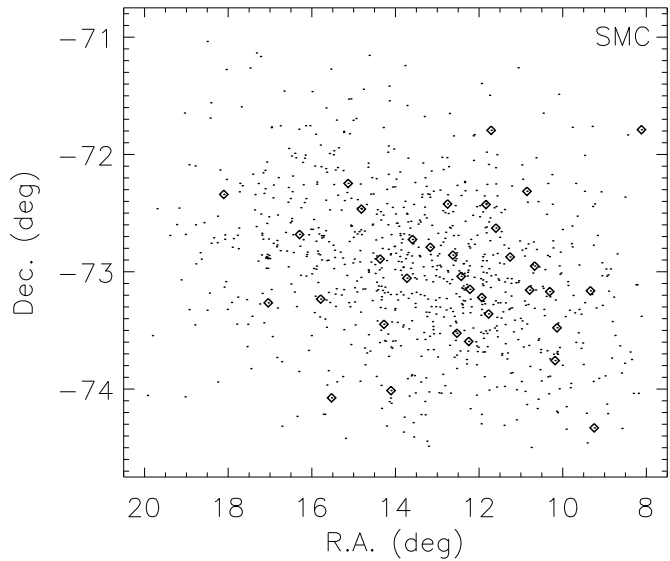}
\caption{Spatial distribution of all SMC AGB stars selected in this paper (dots). The most
luminous with $-8.4 < M_{\rm bol} < -6.0$~mag are represented with diamonds.}
\label{f5}
\end{figure}

The carbon star candidates selected here are distributed within an
elliptical area, with no obvious central concentration (Fig. \ref{f5}).
The brightest sources with $-8.4 < M_{\rm bol} < -6.0$~mag tend to be in the
central area, although they are too few in number to draw a definite
conclusion. Demers et al. (\cite{demers}) found that the carbon stars
are located almost exclusively in and near the disk for the Magellanic
type galaxy NGC 3109. This is consistent with our results in the SMC
and the LMC.

\subsection{Messier 31}
\begin{figure}
\centering
\includegraphics[width=9cm]{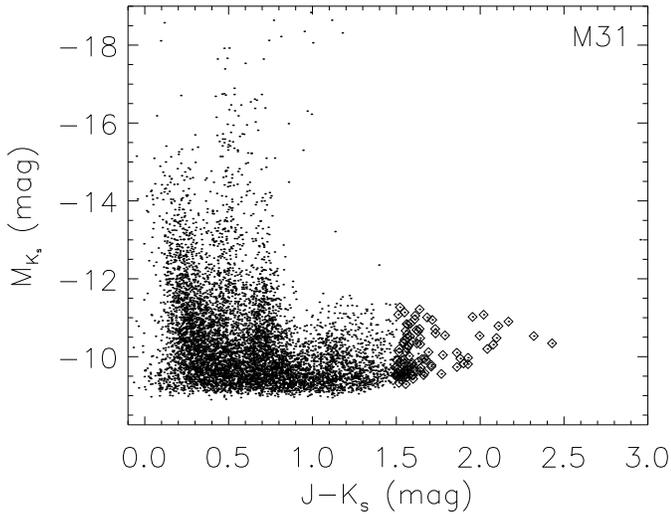}
\caption{Color--magnitude diagram for all M31 stars detected by 2MASS. The
upper part of the {\em red tail} is located at $J-K_s>1.1$ ~mag, whereas the
luminous AGB stars with $-8.4 < M_{\rm bol} < -6.0$~mag are shown with diamonds.}
\label{f6}
\end{figure}

\begin{figure}
\centering
%\includegraphics[width=10cm]{f6.ps}
%%%Yiannis
\setlength{\unitlength}{1cm}
\begin{picture}(9,7)
\put(-1.1,-1){\includegraphics{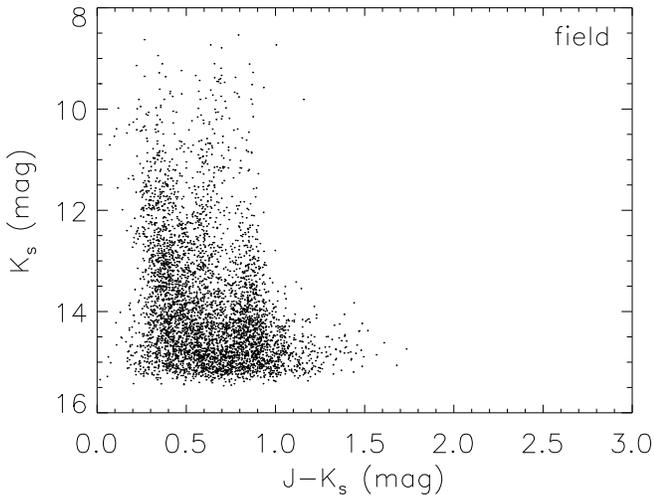}}
\end{picture}
%%%Yiannis
\caption{Color--magnitude diagram for stars detected by 2MASS in a nearby
field.}
\label{f7}
\end{figure}

M31 is a more distant galaxy than the LMC and SMC, with a distance modulus
of $m-M=24.38 \pm 0.05$~mag (Brewer and Richer \cite{brewer2}). The
color--magnitude diagram presented in Fig. \ref{f6} shows that only the
upper part of the {\em red tail} is observable in M31 because of the 2MASS
sensitivity limit. It was assumed that all $J-K_s \ge 1.1$~mag stars
are {\em red tail} stars (see Fig. \ref{f6}).
In that area there are 959 stars with photometric errors less than 0.15~mag.
Since M31 is located at an intermediate galactic latitude, the question of
foreground star contamination needs to be properly addressed.
We retrieved data from a region near M31 (Fig. \ref{f7}), located at $16.0<{\rm RA}<17.5$~deg
and $40.5<{\rm Dec}<42.0$~deg. The color--magnitude diagram for M31
is very populated at $J-K_s>1.1$ ~mag (Fig. \ref{f6}), whereas the neighboring
field (Fig. \ref{f7}) contains very few stars in this part of the diagram,
indicating that the contamination from the Milky Way stellar population
is very low.

For the total reddening correction near M31,
$E_{B-V} \approx 0.23$~mag is adopted (Brewer \& Richer \cite{brewer2}).
Consequently the absorption at $K_s$ is 0.08~mag, assuming again
$A_{K_s}=0.34 \times E_{B-V}$ (Bessel \& Brett \cite{bessel}).
From these relatively low absorption values, we excluded the possibility
that the 959 {\em red tail} stars are early type stars heavily obscured,
although they are located in the spiral structure of M31.

\begin{figure}
\centering
\includegraphics[angle=0,width=9cm]{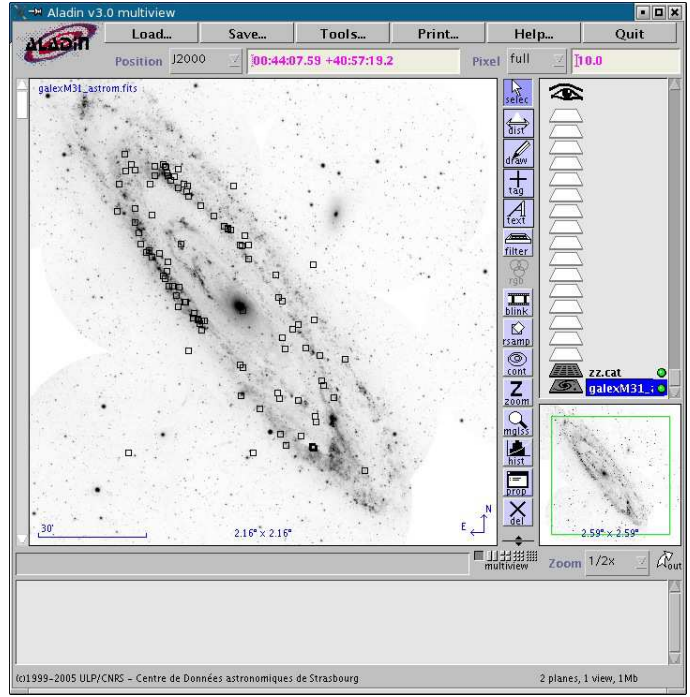}
\caption{The brightest carbon star candidates found in M31 (squares)
overplotted on the GALEX UV image, as processed by Aladin v3.0.}
\label{f8}
\end{figure}

Considering the criteria of Davidge (\cite{davidge2}), only 100 stars are carbon
star candidates. In Fig. \ref{f8} we overplotted the carbon star
candidates found in M31 on the GALEX image. These stars fall on the star
forming regions detected at UV wavelengths, following the familiar M31
ring--structure, 10~kpc away from the galaxy center. This ring has been
detected in most of the large scale maps at radio wavelengths
(Beck \& Gr\"{a}ve \cite{beck}; Brinks \& Shane \cite{brinks}), and
is also traced by star--forming regions (Pellet et al. \cite{pellet};
Devereux et al. \cite{devereux}). This structure can also be noted in
the infrared (Haas et al. \cite{haas}), in the optical via masking
(Walterbos \& Kennicut \cite{walterbos}), and in the distribution of
HII regions (Pellet et al. \cite{pellet}), OB associations (van der
Bergh \cite{van}), HI gas (Sofue \& Kato \cite{sofue}), and other
tracers (see Hodge \cite{hodge}). The spatial distribution of the
carbon star candidates suggests that the velocity dispersion and the
differential galactic rotation have not had enough time to spread these
bright AGB stars, which are among the youngest intermediate mass stars
produced in spiral arms, like the current population of young stars.
All the 100 bright AGB stars detected here are within the magnitude range of
$-8.4 < M_{\rm bol} < -6.0$~mag. They are listed in Table \ref{m31}.

In previous investigations of carbon stars in M31, either the studied areas are
far from the ring of luminous AGB stars presented in this
paper or candidate carbon stars are fainter (Brewer et al.
\cite{brewer1}; Battinelli et al. \cite{battinelli2}; Battinelli et al.
\cite{battinelli1}; Davidge et al. \cite{davidge1}).

\subsection{Messier 33}
\begin{figure}
\centering
\includegraphics[width=9cm]{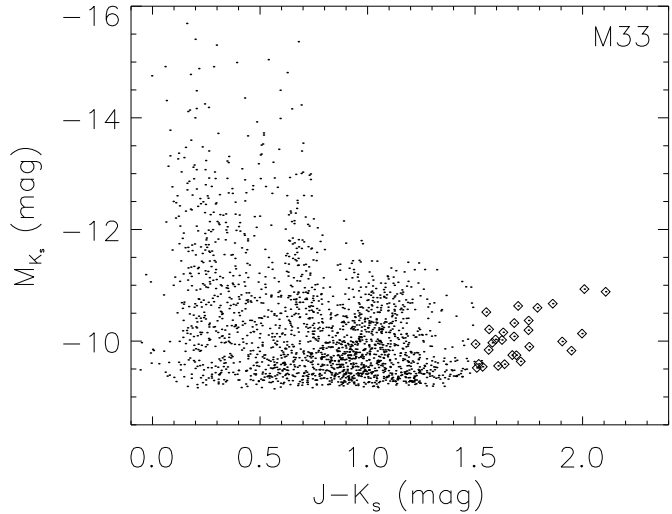}
\caption{Color--magnitude diagram for the stars in M33. The upper part of
the {\em red tail} is located at $J-K > 1.0$~mag, whereas the luminous AGB
stars with $-8.4 < M_{\rm bol} < -6.0$~mag are shown with diamonds.}
\label{f9}
\end{figure}

To complete our survey of the brightest {\em red tail} stars in nearby
galaxies, we took the 2MASS data for M33 and followed the same process as for
the LMC, SMC, and M31. Wilson et al. (\cite{wilson}) derive $E_{B-V}=0.3
\pm 0.1$, including both the Milky Way foreground and M33 internal extinction.
This color excess is translated into $A_{K_s} \approx 0.10$~mag. The distance
modulus for M33 is $m-M=24.52 \pm 0.14 {\rm (statistical)} \pm 0.13 {\rm
(systematic)}$~mag (Lee et al. \cite{lee}), and the {\em red tail} is located
at $J-K_s>1.0$~mag (Fig.  \ref{f9}). Block et al. (\cite{block}) consider
that stars detected by 2MASS with $J-K_s>1.0$~mag cannot be M-stars in the
low-metallicity regions, indicating the presence of {\em red tail} stars.
In this area of the color--magnitude diagram, 916 stars have been found; and
among them, 31 are supposed to be carbon stars, after we adopt the Davidge
(\cite{davidge2}) criteria. We stress that all the 31 carbon
star candidates are very bright with $-8.4 < M_{\rm bol} < -6.0$~mag (see
Table \ref{m33}).

\begin{figure}
\centering
%\includegraphics[width=10cm]{m33.ps}
%%%Yiannis
\setlength{\unitlength}{1cm}
\begin{picture}(10,7)
\put(-1.1,-1){\includegraphics{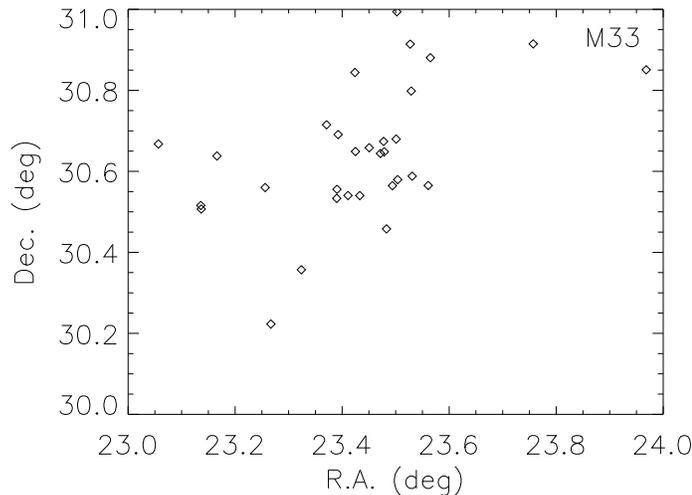}}
\end{picture}
%%%Yiannis
\caption{Spatial distribution of the most
luminous AGB stars with $-8.4 < M_{\rm bol} < -6.0$~mag.}
\label{f10}
\end{figure}

The spatial distribution of these stars (Fig.\ref{f10})  is not correlated with the arcs of
red stars found by Block et al. (\cite{block}) in the disk of M33. The
cross--identification of these 31 carbon star candidates with those of
Rowe et al. (\cite{rowe}) gives no pair within a 1\arcsec\ search radius.
Their study has not revealed the most luminous AGB star population detected
in our paper here, most probably because their observations are in the
optical domain where these stars are not detected.

\section{Discussion and conclusion}
\subsection{Luminosity Functions}
The luminosity function reflects the distribution of stellar masses formed
in a given volume of space for a stellar system or a galaxy. The upper end
of the luminosity function displays the most massive stars of any particular
population. Carbon stars usually have intermediate mass and provide insight
into the star--forming history of these masses.

\begin{figure}
\centering
\includegraphics[angle=0,width=9cm]{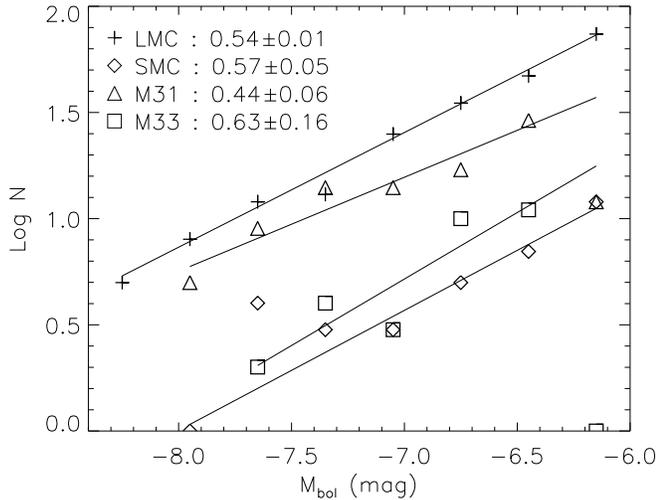}
\caption{Luminosity functions for the luminous carbon and OH/IR candidates
found in the LMC, the SMC, M31, and M33. Numbers in the upper--left corner are
the slopes of the linear regression fits.}
\label{f11}
\end{figure}

\begin{figure}
\centering
\includegraphics[angle=0,width=9cm]{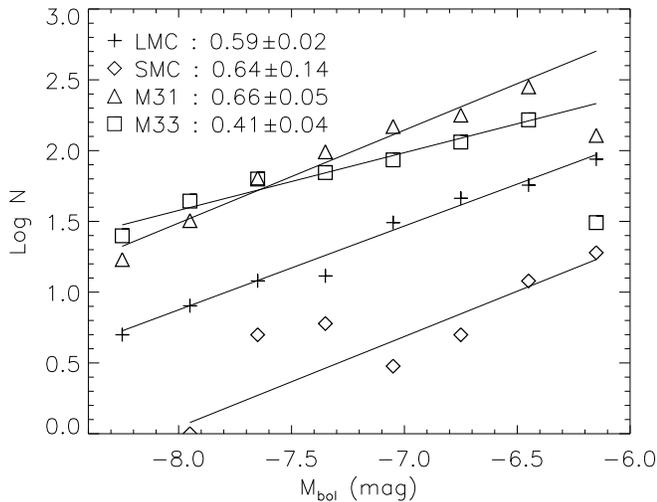}
\caption{Luminosity function for the bright {\em red tail} stars found in the
LMC, the SMC, M31, and M33. Numbers in the upper--left corner are the slopes of the
linear regression fits.}
\label{f12}
\end{figure}

The luminosity function for the bright AGBs in each galaxy is presented in
Fig. \ref{f11} (LMC, SMC, M31, M33). It illustrates that in all four galaxies, luminous carbon and
OH/IR candidates (Egan et al. \cite{egan}) exhibit a similar luminosity function with slopes varying
only slightly and within the errors of the fit. This result suggests that a
similar mass distribution is expected in all four galaxies for the upper part
of the mass function. Groenewegen (\cite{groenewegen}) found that the faint
part of luminosity functions are also similar. Luminosity functions for the
bright {\em red tail} population are presented in Fig. \ref{f12}.
No significant difference is found between the population of bright ($-8.4 < M_{\rm bol} < -6.0$~mag)
AGB {\em red tail} stars ( $H-K_s>0.4$~mag and $J-K_s > 1.5$~mag)
and the whole population of {\em red tail} stars (LMC: $J-K_s > 1.3$~mag and $M_{K_s} < -6.5$~mag,
SMC: $J-K_s > 1.2$~mag and $M_{K_s} < -7.0$~mag, M31: $J-K_s > 1.1$~mag, M33: $J-K_s > 1.0$~mag).

\subsection{The spatial distribution}
In the present work the spatial distribution of the luminous AGB stars
appears to depend on the environment. In the LMC they are distributed within
an ellipse defined by the extent of the old population of the galaxy.
The density is higher along the bar, while the most luminous AGB stars are
concentrated in the central area where the young cluster systems are
located (Kontizas et al., \cite{kontizas2}, Kontizas et al., \cite{kontizas2}).

In M31 the AGB stars are distributed on the spiral features of the galaxy.
Therefore both the LMC and M31 show that the luminous and, consequently, the
most massive AGB stars trace the location of the star--forming regions of their
parent galaxy.
On the other hand, the less massive stars that evolve slowly have the time
to reach the outer parts of the galaxy, as observed for the LMC. To
confirm this behavior in M31, deeper near--infrared data are required.

In the SMC and M33, AGBs are located in an elliptical area along the disk,
without an obvious higher density at star--forming regions, i.e. the center
and the spiral structure. It is difficult to draw any conclusion from this
diagrams since the amount of luminous AGB stars detected in these galaxies is very small.

\subsection{Conclusion}
We searched for the most luminous AGB stars in four nearby galaxies:
the LMC, the SMC, M31, and M33.
Stars were selected on the {\em red tail} of the color--magnitude
diagram ($M_{K_s}$, $J-K_s$) with $J-K_s >1.5$~mag and $H-K_s >0.4$~mag.
Known optical catalogues of carbon stars in the SMC and LMC do not include
these luminous AGB stars, since they were constructed from observations
at $B$ and $R$ wavelengths; therefore,
the catalogues of luminous AGB stars compiled in this work are a
valuable complement to existing data.

Using the same range in $M_{\rm bol}$ ($-8.4 < M_{\rm bol} < -6.0$~mag) for the
four galaxies, we analyzed the spatial distribution of these stars.
We found that they follow the spiral ring of M31 and are mostly
located in the central region of the LMC, where all the young population is
concentrated. In the SMC and in M33, their number is too small to draw reliable
conclusions about their location. Finally the luminosity functions of the {\em red tail}
stars and the bright AGB stars, display similar slopes
within the same bolometric magnitude range.

\section{Acknowledgments}
The authors would like to thank the CDS and the VO for valuable support. P.
Tsalmantza and M. Kontizas would like to thank the E.L.K.E. of the
University of Athens and the Greek General Secretariat of Research
\& Technology. Special thanks go to the University of Strasbourg for
financial support and the hospitality of E. \& M. Kontizas.
This publication makes use of data products from the Two Micron All Sky Survey,
which is a joint project of the University of Massachusetts and the Infrared
Processing and Analysis Center/California Institute of Technology, funded by the
National Aeronautics and Space Administration and the National Science Foundation.
This research made use of the VizieR catalogue access tool and Aladin, CDS,
Strasbourg, France. Finally, but not least, we would like to thank the unknown referee for
very constructive and stimulating comments.

\clearpage
\onecolumn
\appendix
\section{Tables}

\vspace{1cm}
\begin{center}
\topcaption{Very luminous AGB stars (carbon star candidates) in the LMC}
\label{lmc}
\tablefirsthead{
  \hline
  \hline
  2MASS ID & RA   & Dec   & $M_{K}$ & $M_{\rm bol}$ & $J-K_s$ \\
           &(deg) & (deg) & (mag)   & (mag) & (mag)\\
  \hline}
  \tablehead{
  \hspace{-1cm}table continuation\\
    \hline
  2MASS ID & RA   & Dec   & $M_{K}$ & $M_{\rm bol}$ & $J-K_s$ \\
           &(deg) & (deg) & (mag)   & (mag) & (mag)\\
  \hline}
  \tabletail{\hline}
\begin{supertabular}{lrrrrr}
2MASS J06103966-6655227 &  92.665289 & -66.922974 &  -6.639 &  -6.323 &  
4.10\\
2MASS J06022433-6619263 &  90.601393 & -66.323982 &  -7.723 &  -7.073 &  
3.98\\
2MASS J05294752-7609256 &  82.448038 & -76.157127 &  -7.677 &  -6.752 &  
3.88\\
2MASS J04564410-7442267 &  74.183760 & -74.707420 &  -7.638 &  -6.831 &  
3.92\\
2MASS J06011982-6525021 &  90.332610 & -65.417274 &  -7.844 &  -6.238 &  
3.59\\
2MASS J06055819-6522162 &  91.492478 & -65.371185 &  -7.392 &  -6.798 &  
4.00\\
2MASS J06094920-6517197 &  92.455038 & -65.288811 &  -6.541 &  -6.496 &  
4.20\\
2MASS J05412924-7336038 &  85.371848 & -73.601082 &  -7.331 &  -6.190 &  
3.79\\
2MASS J05312948-7326331 &  82.872862 & -73.442551 &  -7.010 &  -6.314 &  
3.96\\
2MASS J05074394-7312521 &  76.933107 & -73.214478 &  -7.318 &  -7.460 &  
4.26\\
2MASS J05335481-7312143 &  83.478390 & -73.203995 &  -6.979 &  -6.857 &  
4.17\\
2MASS J05312710-7307148 &  82.862936 & -73.120804 &  -7.485 &  -6.878 &  
4.00\\
2MASS J05455697-7254291 &  86.487392 & -72.908104 &  -7.530 &  -6.134 &  
3.68\\
2MASS J05203823-7251062 &  80.159310 & -72.851723 &  -7.644 &  -6.709 &  
3.87\\
2MASS J05280053-7250120 &  82.002241 & -72.836685 &  -7.038 &  -6.050 &  
3.85\\
2MASS J05474046-7248485 &  86.918596 & -72.813484 &  -6.843 &  -8.045 &  
4.58\\
2MASS J05422827-7225333 &  85.617810 & -72.425926 &  -7.652 &  -6.173 &  
3.64\\
2MASS J05480721-7221516 &  87.030047 & -72.364342 &  -6.636 &  -6.315 &  
4.10\\
2MASS J05085905-7209546 &  77.246083 & -72.165176 &  -7.452 &  -7.116 &  
4.10\\
2MASS J05152480-7154493 &  78.853353 & -71.913696 &  -6.927 &  -7.090 &  
4.26\\
2MASS J05200955-7152132 &  80.039798 & -71.870346 &  -6.927 &  -6.793 &  
4.17\\
2MASS J05484335-7151165 &  87.180628 & -71.854607 &  -7.178 &  -6.536 &  
3.98\\
2MASS J05242503-7149020 &  81.104331 & -71.817238 &  -7.547 &  -6.862 &  
3.97\\
2MASS J05423112-7146588 &  85.629679 & -71.783005 &  -7.226 &  -7.522 &  
4.31\\
2MASS J05500676-7146026 &  87.528169 & -71.767410 &  -7.536 &  -6.950 &  
4.01\\
2MASS J06190234-7321026 &  94.759790 & -73.350746 &  -6.923 &  -7.918 &  
4.52\\
2MASS J05024447-7139323 &  75.685300 & -71.658974 &  -7.097 &  -7.602 &  
4.37\\
2MASS J05490888-7132069 &  87.287000 & -71.535255 &  -9.085 &  -6.078 &  
2.68\\
2MASS J05465088-7128033 &  86.712036 & -71.467598 &  -6.871 &  -7.805 &  
4.50\\
2MASS J05300380-7124361 &  82.515852 & -71.410049 &  -6.690 &  -7.112 &  
4.35\\
2MASS J05313006-7121448 &  82.875254 & -71.362465 &  -7.869 &  -6.774 &  
3.81\\
2MASS J05013788-7121123 &  75.407859 & -71.353432 &  -7.305 &  -8.313 &  
4.53\\
2MASS J05350379-7113576 &  83.765795 & -71.232681 &  -7.885 &  -6.174 &  
3.54\\
2MASS J05270395-7108591 &  81.766488 & -71.149750 &  -7.898 &  -7.250 &  
3.98\\
2MASS J06241389-7229196 &  96.057905 & -72.488785 &  -9.307 &  -6.005 &  
2.22\\
2MASS J05251951-7104027 &  81.331302 & -71.067429 &  -7.859 &  -7.861 &  
4.21\\
2MASS J06092546-7226037 &  92.356121 & -72.434364 &  -7.846 &  -6.333 &  
3.63\\
2MASS J05284817-7102289 &  82.200709 & -71.041374 &  -8.713 &  -6.151 &  
3.06\\
2MASS J05345289-7100229 &  83.720388 & -71.006363 &  -8.853 &  -6.035 &  
2.86\\
2MASS J05535326-7059475 &  88.471917 & -70.996529 &  -7.462 &  -6.149 &  
3.72\\
2MASS J05482201-7058476 &  87.091717 & -70.979904 &  -7.802 &  -6.577 &  
3.75\\
2MASS J05202105-7057545 &  80.087720 & -70.965149 &  -7.620 &  -7.114 &  
4.03\\
2MASS J05061466-7056427 &  76.561102 & -70.945198 &  -6.701 &  -6.724 &  
4.22\\
2MASS J05381238-7056174 &  84.551612 & -70.938194 &  -6.967 &  -6.109 &  
3.90\\
2MASS J05221654-7053460 &  80.568919 & -70.896126 &  -7.635 &  -6.332 &  
3.72\\
2MASS J04532183-7051449 &  73.340999 & -70.862488 &  -7.933 &  -8.105 &  
4.27\\
2MASS J04343118-7049491 &  68.629954 & -70.830322 &  -6.548 &  -6.278 &  
4.12\\
2MASS J05465510-7047254 &  86.729615 & -70.790390 &  -7.808 &  -7.017 &  
3.93\\
2MASS J06054883-7146462 &  91.453467 & -71.779518 &  -7.313 &  -6.865 &  
4.06\\
2MASS J05415782-7042418 &  85.490917 & -70.711617 &  -7.710 &  -6.627 &  
3.81\\
2MASS J04573049-7036575 &  74.377079 & -70.615990 &  -8.211 &  -6.302 &  
3.44\\
2MASS J05241094-7036210 &  81.045601 & -70.605843 &  -8.243 &  -6.430 &  
3.49\\
2MASS J05281148-7033586 &  82.047840 & -70.566299 &  -6.704 &  -6.192 &  
4.03\\
2MASS J05530336-7033172 &  88.264027 & -70.554779 &  -9.082 &  -7.163 &  
3.43\\
2MASS J05464319-7033154 &  86.679960 & -70.554283 &  -7.032 &  -6.325 &  
3.96\\
2MASS J05350751-7032195 &  83.781320 & -70.538757 &  -7.798 &  -6.337 &  
3.65\\
2MASS J05343646-7032135 &  83.651938 & -70.537094 &  -6.577 &  -6.176 &  
4.07\\
2MASS J05103253-7030497 &  77.635542 & -70.513832 &  -8.302 &  -7.103 &  
3.77\\
2MASS J05424425-7030208 &  85.684407 & -70.505783 &  -7.175 &  -6.434 &  
3.95\\
2MASS J05032665-7029090 &  75.861058 & -70.485855 &  -7.256 &  -6.203 &  
3.83\\
2MASS J05165826-7029085 &  79.242791 & -70.485703 &  -8.072 &  -6.113 &  
3.41\\
2MASS J05152170-7027315 &  78.840457 & -70.458771 &  -7.904 &  -6.295 &  
3.59\\
2MASS J06162403-7015127 &  94.100149 & -70.253555 &  -6.772 &  -6.047 &  
3.95\\
2MASS J05273669-7024036 &  81.902897 & -70.401024 &  -7.186 &  -6.890 &  
4.11\\
2MASS J05231239-7022045 &  80.801638 & -70.367935 &  -7.657 &  -6.528 &  
3.79\\
2MASS J05120517-7021582 &  78.021575 & -70.366173 &  -7.155 &  -6.945 &  
4.14\\
2MASS J06121430-6943119 &  93.059601 & -69.719978 &  -6.535 &  -6.704 &  
4.27\\
2MASS J05060423-7016513 &  76.517648 & -70.280930 & -10.087 &  -6.823 &  
1.64\\
2MASS J06094093-6847002 &  92.420551 & -68.783401 &  -8.081 &  -7.207 &  
3.90\\
2MASS J05135752-7014104 &  78.489704 & -70.236237 &  -7.903 &  -6.689 &  
3.76\\
2MASS J06083184-6816353 &  92.132673 & -68.276482 &  -6.743 &  -6.633 &  
4.17\\
2MASS J05332717-7009516 &  83.363228 & -70.164352 &  -6.893 &  -6.035 &  
3.90\\
2MASS J04334368-7009504 &  68.432034 & -70.164024 &  -6.696 &  -6.089 &  
4.00\\
2MASS J05390173-7008429 &  84.757222 & -70.145256 &  -7.737 &  -7.490 &  
4.13\\
2MASS J05253612-7007235 &  81.400528 & -70.123215 &  -8.346 &  -6.029 &  
3.21\\
2MASS J05444714-7007047 &  86.196438 & -70.117973 &  -7.499 &  -6.094 &  
3.68\\
2MASS J05381127-7006088 &  84.546982 & -70.102470 &  -9.649 &  -6.417 &  
1.58\\
2MASS J05470053-7003170 &  86.752248 & -70.054733 &  -7.347 &  -7.370 &  
4.22\\
2MASS J05481114-7000167 &  87.046428 & -70.004646 &  -6.551 &  -6.092 &  
4.05\\
2MASS J05552103-7000030 &  88.837643 & -70.000847 &  -9.463 &  -6.573 &  
2.80\\
2MASS J05351559-6958443 &  83.814974 & -69.978996 &  -7.858 &  -6.365 &  
3.64\\
2MASS J04491008-6958048 &  72.292016 & -69.968002 &  -7.674 &  -6.033 &  
3.57\\
2MASS J05095999-6956097 &  77.499982 & -69.936028 &  -8.574 &  -6.449 &  
3.32\\
2MASS J05250941-6955407 &  81.289231 & -69.927994 &  -8.887 &  -6.116 &  
2.90\\
2MASS J05455416-6955172 &  86.475667 & -69.921448 &  -7.997 &  -6.108 &  
3.45\\
2MASS J04402848-6955135 &  70.118687 & -69.920441 &  -9.897 &  -6.560 &  
1.90\\
2MASS J05300859-6954496 &  82.535810 & -69.913795 &  -8.537 &  -6.263 &  
3.24\\
2MASS J05162261-6954181 &  79.094245 & -69.905037 &  -7.687 &  -6.250 &  
3.66\\
2MASS J04491848-6953145 &  72.327023 & -69.887367 &  -9.159 &  -6.764 &  
3.17\\
2MASS J05403606-6952498 &  85.150270 & -69.880508 &  -8.233 &  -7.906 &  
4.10\\
2MASS J05350354-6952454 &  83.764785 & -69.879295 &  -7.936 &  -7.505 &  
4.06\\
2MASS J05352686-6952279 &  83.861930 & -69.874420 &  -7.536 &  -6.787 &  
3.94\\
2MASS J04485265-6951334 &  72.219410 & -69.859299 &  -7.282 &  -8.246 &  
4.51\\
2MASS J05251232-6950377 &  81.301366 & -69.843819 &  -8.325 &  -6.364 &  
3.41\\
2MASS J05254237-6947129 &  81.426567 & -69.786942 &  -7.170 &  -6.132 &  
3.83\\
2MASS J05245135-6947042 &  81.213967 & -69.784508 &  -8.109 &  -6.079 &  
3.38\\
2MASS J05153398-6945590 &  78.891600 & -69.766396 &  -7.161 &  -7.737 &  
4.40\\
2MASS J05481349-6945115 &  87.056240 & -69.753204 &  -7.229 &  -7.208 &  
4.20\\
2MASS J05282342-6943387 &  82.097615 & -69.727425 &  -6.849 &  -7.552 &  
4.43\\
2MASS J05220985-6943200 &  80.541050 & -69.722244 &  -7.934 &  -6.580 &  
3.70\\
2MASS J05324622-6942497 &  83.192614 & -69.713829 &  -8.458 &  -6.152 &  
3.22\\
2MASS J05175887-6939231 &  79.495332 & -69.656433 &  -9.485 &  -6.281 &  
1.53\\
2MASS J05161189-6938596 &  79.049581 & -69.649895 &  -7.222 &  -6.224 &  
3.85\\
2MASS J04502338-6937567 &  72.597436 & -69.632431 &  -7.458 &  -7.046 &  
4.07\\
2MASS J05103243-6936542 &  77.635141 & -69.615059 &  -7.572 &  -6.414 &  
3.78\\
2MASS J05105680-6935303 &  77.736668 & -69.591759 &  -7.584 &  -7.638 &  
4.23\\
2MASS J05240917-6935112 &  81.038249 & -69.586456 &  -9.071 &  -6.111 &  
2.73\\
2MASS J05455384-6931233 &  86.474335 & -69.523140 &  -9.492 &  -6.263 &  
1.57\\
2MASS J04283018-6930502 &  67.125786 & -69.513962 &  -6.732 &  -7.409 &  
4.43\\
2MASS J05031358-6930326 &  75.806610 & -69.509079 &  -7.471 &  -6.030 &  
3.66\\
2MASS J05271416-6929112 &  81.809034 & -69.486465 &  -6.920 &  -6.140 &  
3.93\\
2MASS J05273189-6928353 &  81.882880 & -69.476494 &  -6.614 &  -6.521 &  
4.18\\
2MASS J05070471-6928053 &  76.769657 & -69.468163 &  -7.720 &  -6.261 &  
3.65\\
2MASS J05163771-6927142 &  79.157138 & -69.453964 &  -7.160 &  -7.821 &  
4.42\\
2MASS J04553354-6924593 &  73.889761 & -69.416473 & -10.266 &  -7.034 &  
1.58\\
2MASS J05401333-6922464 &  85.055542 & -69.379578 &  -9.988 &  -6.670 &  
1.79\\
2MASS J05123400-6922086 &  78.141669 & -69.369057 &  -7.455 &  -6.432 &  
3.84\\
2MASS J04584149-6921314 &  74.672881 & -69.358727 &  -6.960 &  -7.074 &  
4.25\\
2MASS J04473884-6921170 &  71.911845 & -69.354729 &  -8.041 &  -6.411 &  
3.58\\
2MASS J06145902-6821262 &  93.745929 & -68.357285 &  -7.088 &  -6.035 &  
3.83\\
2MASS J05383397-6920317 &  84.641553 & -69.342140 & -10.082 &  -6.817 &  
1.64\\
2MASS J05171633-6920298 &  79.318059 & -69.341614 &  -8.005 &  -6.071 &  
3.43\\
2MASS J05325547-6920266 &  83.231141 & -69.340736 &  -6.864 &  -6.452 &  
4.07\\
2MASS J05170066-6919304 &  79.252753 & -69.325127 &  -7.770 &  -7.598 &  
4.15\\
2MASS J05042995-6919235 &  76.124802 & -69.323212 &  -7.291 &  -7.154 &  
4.16\\
2MASS J05085218-6916337 &  77.217443 & -69.276039 &  -7.114 &  -6.088 &  
3.84\\
2MASS J05123206-6915404 &  78.133621 & -69.261230 &  -7.602 &  -6.697 &  
3.88\\
2MASS J05364101-6914064 &  84.170893 & -69.235138 &  -9.293 &  -6.038 &  
1.62\\
2MASS J04524566-6911494 &  73.190282 & -69.197075 &  -8.631 &  -7.824 &  
3.92\\
2MASS J04430510-6909128 &  70.771287 & -69.153557 &  -7.530 &  -7.206 &  
4.10\\
2MASS J05303638-6907099 &  82.651612 & -69.119438 &  -9.715 &  -6.516 &  
1.52\\
2MASS J05292199-6906584 &  82.341629 & -69.116226 &  -7.729 &  -6.970 &  
3.94\\
2MASS J05101432-6906103 &  77.559695 & -69.102882 &  -7.222 &  -6.426 &  
3.93\\
2MASS J05102724-6904532 &  77.613541 & -69.081467 &  -6.994 &  -6.619 &  
4.08\\
2MASS J05023218-6904377 &  75.634084 & -69.077148 &  -6.885 &  -7.134 &  
4.29\\
2MASS J05384137-6903540 &  84.672401 & -69.065002 &  -7.076 &  -6.948 &  
4.17\\
2MASS J05202143-6858479 &  80.089325 & -68.979973 &  -6.628 &  -6.105 &  
4.03\\
2MASS J04554180-6857227 &  73.924175 & -68.956329 &  -7.858 &  -6.827 &  
3.83\\
2MASS J05145597-6856464 &  78.733248 & -68.946243 &  -6.756 &  -6.500 &  
4.12\\
2MASS J05224292-6855288 &  80.678873 & -68.924683 &  -7.653 &  -7.455 &  
4.14\\
2MASS J05000498-6853553 &  75.020789 & -68.898720 &  -9.182 &  -6.136 &  
2.64\\
2MASS J04563215-6852510 &  74.133960 & -68.880844 &  -7.466 &  -7.790 &  
4.32\\
2MASS J05152176-6849019 &  78.840680 & -68.817215 &  -7.409 &  -7.039 &  
4.08\\
2MASS J05215630-6847225 &  80.484594 & -68.789589 &  -6.969 &  -7.818 &  
4.48\\
2MASS J05102834-6844313 &  77.618120 & -68.742050 &  -7.244 &  -6.511 &  
3.95\\
2MASS J04511548-6841403 &  72.814508 & -68.694534 &  -7.498 &  -6.430 &  
3.82\\
2MASS J05151758-6841198 &  78.823286 & -68.688835 &  -6.811 &  -6.433 &  
4.08\\
2MASS J04390199-6836322 &  69.758300 & -68.608963 &  -7.050 &  -6.714 &  
4.10\\
2MASS J04522647-6834374 &  73.110316 & -68.577080 &  -9.619 &  -6.350 &  
1.65\\
2MASS J05565154-6827267 &  89.214751 & -68.457420 &  -9.982 &  -6.665 &  
1.79\\
2MASS J05474884-6825336 &  86.953524 & -68.426003 &  -8.143 &  -6.340 &  
3.49\\
2MASS J04471609-6824256 &  71.817052 & -68.407120 &  -7.936 &  -6.677 &  
3.74\\
2MASS J04552049-6822391 &  73.835412 & -68.377541 &  -8.865 &  -6.074 &  
2.88\\
2MASS J04554249-6816542 &  73.927067 & -68.281723 &  -7.408 &  -6.360 &  
3.83\\
2MASS J04585546-6813062 &  74.731110 & -68.218407 &  -7.537 &  -6.276 &  
3.74\\
2MASS J05325618-6812487 &  83.234091 & -68.213539 &  -9.268 &  -6.358 &  
2.78\\
2MASS J04544534-6804146 &  73.688926 & -68.070747 &  -8.110 &  -6.008 &  
3.34\\
2MASS J05081566-6800459 &  77.065279 & -68.012764 &  -6.536 &  -6.002 &  
4.02\\
2MASS J05213755-6752393 &  80.406488 & -67.877594 &  -7.261 &  -6.149 &  
3.80\\
2MASS J04594477-6752208 &  74.936568 & -67.872452 &  -8.599 &  -6.264 &  
3.20\\
2MASS J05111047-6752105 &  77.793643 & -67.869591 &  -6.900 &  -8.385 &  
4.66\\
2MASS J05502604-6749462 &  87.608526 & -67.829506 &  -7.659 &  -6.966 &  
3.97\\
2MASS J05214756-6749118 &  80.448169 & -67.819962 &  -8.728 &  -6.123 &  
3.03\\
2MASS J05100440-6745501 &  77.518363 & -67.763939 &  -8.521 &  -6.009 &  
3.09\\
2MASS J05150378-6744142 &  78.765788 & -67.737289 &  -7.545 &  -6.310 &  
3.75\\
2MASS J04553202-6742310 &  73.883426 & -67.708626 &  -7.358 &  -7.068 &  
4.11\\
2MASS J05195418-6739420 &  79.975765 & -67.661674 &  -7.001 &  -6.063 &  
3.87\\
2MASS J04333653-6739329 &  68.402242 & -67.659157 &  -6.798 &  -7.252 &  
4.36\\
2MASS J05065934-6734453 &  76.747273 & -67.579254 &  -8.096 &  -6.033 &  
3.36\\
2MASS J04521072-6734329 &  73.044681 & -67.575813 &  -6.701 &  -6.414 &  
4.11\\
2MASS J05503932-6731504 &  87.663860 & -67.530685 &  -7.070 &  -7.581 &  
4.38\\
2MASS J05342976-6730257 &  83.624008 & -67.507164 &  -7.839 &  -6.198 &  
3.57\\
2MASS J05054710-6727243 &  76.446275 & -67.456772 &  -7.404 &  -8.260 &  
4.48\\
2MASS J05135297-6726548 &  78.470730 & -67.448563 & -10.128 &  -6.799 &  
1.85\\
2MASS J05081617-6723506 &  77.067413 & -67.397408 &  -7.179 &  -6.069 &  
3.80\\
2MASS J04572171-6721280 &  74.340464 & -67.357780 &  -7.387 &  -6.021 &  
3.69\\
2MASS J05281556-6720188 &  82.064839 & -67.338562 &  -7.590 &  -6.745 &  
3.91\\
2MASS J05283913-6718089 &  82.163055 & -67.302483 &  -6.798 &  -6.499 &  
4.11\\
2MASS J04542493-6709344 &  73.603899 & -67.159569 &  -7.291 &  -7.640 &  
4.32\\
2MASS J05001899-6707580 &  75.079159 & -67.132782 &  -9.267 &  -6.234 &  
2.66\\
2MASS J05323716-6706564 &  83.154849 & -67.115669 &  -7.943 &  -6.586 &  
3.70\\
2MASS J04352409-6656493 &  68.850409 & -66.947029 &  -6.874 &  -6.835 &  
4.20\\
2MASS J04545344-6651375 &  73.722698 & -66.860435 &  -8.841 &  -6.201 &  
3.00\\
2MASS J05113864-6651098 &  77.911028 & -66.852730 &  -7.853 &  -6.151 &  
3.54\\
2MASS J04253216-6647163 &  66.384027 & -66.787888 &  -7.505 &  -6.941 &  
4.01\\
2MASS J05062118-6643161 &  76.588262 & -66.721153 &  -6.932 &  -7.655 &  
4.44\\
2MASS J06085255-6727472 &  92.218968 & -67.463135 &  -7.407 &  -6.352 &  
3.82\\
2MASS J05365212-6631004 &  84.217194 & -66.516792 &  -7.317 &  -6.745 &  
4.01\\
2MASS J04433797-6628577 &  70.908249 & -66.482704 &  -7.934 &  -6.141 &  
3.50\\
2MASS J05383208-6615382 &  84.633676 & -66.260620 &  -8.882 &  -6.231 &  
2.99\\
2MASS J05311311-6609411 &  82.804646 & -66.161438 &  -7.045 &  -6.932 &  
4.17\\
2MASS J04512141-6609328 &  72.839233 & -66.159134 &  -7.513 &  -6.138 &  
3.69\\
2MASS J05385059-6609200 &  84.710828 & -66.155571 &  -8.543 &  -6.284 &  
3.25\\
2MASS J05320136-6603279 &  83.005668 & -66.057762 &  -7.634 &  -6.406 &  
3.75\\
2MASS J05100176-6551565 &  77.507357 & -65.865715 &  -7.153 &  -6.289 &  
3.90\\
2MASS J04555834-6544052 &  73.993113 & -65.734795 &  -9.244 &  -6.019 &  
2.39\\
2MASS J05225895-6544029 &  80.745639 & -65.734154 &  -7.436 &  -7.157 &  
4.12\\
2MASS J04444769-6537072 &  71.198747 & -65.618690 &  -7.080 &  -7.329 &  
4.29\\
2MASS J05280808-6535201 &  82.033687 & -65.588928 &  -8.207 &  -6.633 &  
3.60\\
2MASS J05242363-6532107 &  81.098476 & -65.536324 &  -7.462 &  -6.007 &  
3.66\\
2MASS J04470329-6524563 &  71.763718 & -65.415649 &  -6.761 &  -7.474 &  
4.44\\
2MASS J05524672-6520071 &  88.194668 & -65.335320 &  -6.935 &  -7.064 &  
4.25\\
2MASS J05353972-6519564 &  83.915505 & -65.332336 &  -6.717 &  -6.018 &  
3.96\\
2MASS J05121594-6518431 &  78.066420 & -65.311981 &  -7.344 &  -6.286 &  
3.82\\
2MASS J05275737-6517363 &  81.989046 & -65.293442 &  -7.775 &  -6.799 &  
3.86\\
2MASS J05295356-6514567 &  82.473182 & -65.249100 &  -7.415 &  -6.485 &  
3.87\\
2MASS J05403844-6457245 &  85.160171 & -64.956818 &  -7.480 &  -7.262 &  
4.14\\
2MASS J05164004-6439127 &  79.166850 & -64.653542 &  -7.921 &  -7.117 &  
3.92\\
2MASS J05501757-6428545 &  87.573216 & -64.481812 &  -8.407 &  -6.115 &  
3.23\\
2MASS J05340710-6416479 &  83.529602 & -64.279976 &  -7.959 &  -6.297 &  
3.56\\
2MASS J05123875-6412136 &  78.161467 & -64.203796 &  -8.960 &  -6.589 &  
3.18\\
2MASS J05310498-6408306 &  82.770769 & -64.141853 &  -8.952 &  -6.505 &  
3.13\\
2MASS J05255025-6400090 &  81.459387 & -64.002502 &  -8.114 &  -6.064 &  
3.37\\
2MASS J05273859-6330540 &  81.910813 & -63.515015 &  -7.055 &  -7.599 &  
4.39\\
2MASS J05152379-6321528 &  78.849158 & -63.364693 &  -6.960 &  -6.736 &  
4.13\\
2MASS J07193710-6559010 & 109.904590 & -65.983627 &  -7.339 &  -6.559 &  
3.93\\
\end{supertabular}
\end{center}

\vspace{1cm}
\begin{center}
\topcaption{Very luminous AGB stars (carbon star candidates) in the SMC}
\label{smc}
\tablefirsthead{
  \hline
  \hline
  2MASS ID & RA   & Dec   & $M_{K}$ & $M_{\rm bol}$ & $J-K_s$ \\
           &(deg) & (deg) & (mag)   & (mag) & (mag)\\
  \hline}
  \tablehead{
  \hspace{-1cm}table continuation\\
    \hline
  2MASS ID & RA   & Dec   & $M_{K}$ & $M_{\rm bol}$ & $J-K_s$ \\
           &(deg) & (deg) & (mag)   & (mag) & (mag)\\
  \hline}
  \tabletail{\hline}
\begin{supertabular}{lrrrrr}
2MASS J00365957-7419503 &   9.248213 & -74.330643 &  -9.886 &  -6.564 &  
2.14\\
2MASS J01020652-7404291 &  15.527175 & -74.074753 &  -8.164 &  -6.114 &  
3.37\\
2MASS J00562548-7400468 &  14.106208 & -74.013023 &  -7.764 &  -6.071 &  
3.55\\
2MASS J00404385-7345244 &  10.182728 & -73.756790 &  -7.558 &  -7.911 &  
4.33\\
2MASS J00485947-7335387 &  12.247813 & -73.594093 &  -7.889 &  -7.729 &  
4.16\\
2MASS J00500719-7331251 &  12.529961 & -73.523666 & -10.381 &  -7.437 &  
2.75\\
2MASS J00403293-7328399 &  10.137230 & -73.477760 &  -7.453 &  -6.860 &  
4.00\\
2MASS J00570590-7326518 &  14.274601 & -73.447746 &  -8.337 &  -6.363 &  
3.41\\
2MASS J00470552-7321330 &  11.773011 & -73.359169 &  -8.700 &  -6.173 &  
3.08\\
2MASS J01081031-7315524 &  17.042983 & -73.264565 &  -9.432 &  -6.208 &  
1.56\\
2MASS J01030900-7313583 &  15.787516 & -73.232887 &  -7.260 &  -7.402 &  
4.26\\
2MASS J00474454-7313072 &  11.935618 & -73.218681 &  -8.607 &  -6.685 &  
3.43\\
2MASS J00411442-7310091 &  10.310122 & -73.169220 &  -7.046 &  -7.133 &  
4.24\\
2MASS J00372080-7309447 &   9.336667 & -73.162430 &  -7.173 &  -6.981 &  
4.15\\
2MASS J00430955-7309223 &  10.789804 & -73.156219 &  -8.230 &  -6.134 &  
3.34\\
2MASS J00485250-7308568 &  12.218778 & -73.149124 &  -7.905 &  -6.197 &  
3.54\\
2MASS J00545410-7303181 &  13.725440 & -73.055054 &  -8.235 &  -6.485 &  
3.52\\
2MASS J00494271-7302207 &  12.427986 & -73.039093 &  -7.144 &  -6.136 &  
3.84\\
2MASS J00424090-7257057 &  10.670455 & -72.951599 &  -8.857 &  -6.236 &  
3.01\\
2MASS J00572770-7253279 &  14.365420 & -72.891106 &  -7.159 &  -6.634 &  
4.03\\
2MASS J00450214-7252243 &  11.258941 & -72.873428 &  -7.828 &  -7.414 &  
4.07\\
2MASS J00503062-7251298 &  12.627616 & -72.858299 & -10.237 &  -7.044 &  
1.51\\
2MASS J00524017-7247276 &  13.167399 & -72.791023 &  -9.174 &  -6.085 &  
2.59\\
2MASS J00542228-7243296 &  13.592872 & -72.724907 &  -7.307 &  -7.554 &  
4.29\\
2MASS J01051049-7240563 &  16.293740 & -72.682327 &  -7.115 &  -6.788 &  
4.10\\
2MASS J00462352-7237397 &  11.598031 & -72.627708 &  -7.306 &  -6.356 &  
3.87\\
2MASS J00591577-7227546 &  14.815710 & -72.465179 &  -8.049 &  -6.434 &  
3.58\\
2MASS J00472059-7225360 &  11.835832 & -72.426689 &  -7.010 &  -7.561 &  
4.39\\
2MASS J00510074-7225185 &  12.753117 & -72.421829 &  -8.031 &  -6.679 &  
3.70\\
2MASS J01122496-7220248 &  18.104013 & -72.340225 &  -7.176 &  -7.542 &  
4.33\\
2MASS J00432514-7218511 &  10.854768 & -72.314217 &  -9.343 &  -6.026 &  
2.17\\
2MASS J01003166-7214489 &  15.131924 & -72.246941 &  -8.041 &  -6.111 &  
3.43\\
2MASS J00465078-7147393 &  11.711623 & -71.794250 &  -7.560 &  -6.015 &  
3.61\\
2MASS J00322809-7147207 &   8.117061 & -71.789093 &  -7.673 &  -6.391 &  
3.73\\
\end{supertabular}
\end{center}

\vspace{1cm}
\begin{center}
\topcaption{Very luminous AGB stars (carbon star candidates) in M31}
\label{m31}
\tablefirsthead{
  \hline
  \hline
  2MASS ID & RA   & Dec   & $M_{K}$ & $M_{\rm bol}$ & $J-K_s$\\
           &(deg) & (deg) & (mag)   & (mag) & (mag)\\
  \hline}
  \tablehead{
  \hspace{-1cm}table continuation\\
    \hline
  2MASS ID & RA   & Dec   & $M_{K}$ & $M_{\rm bol}$ & $J-K_s$\\
           &(deg) & (deg) & (mag)   & (mag) & (mag)\\
  \hline}
  \tabletail{\hline}
\begin{supertabular}{lrrrrr}
2MASS J00394632+4030280 &   9.943011 &  40.507805 & -10.339 &  -7.083 &  
1.62\\
2MASS J00422751+4035524 &  10.614644 &  40.597908 &  -9.834 &  -6.591 &  
1.60\\
2MASS J00452772+4035589 &  11.365515 &  40.599712 &  -9.514 &  -6.325 &  
1.50\\
2MASS J00405951+4036490 &  10.247993 &  40.613636 & -11.009 &  -7.727 &  
1.68\\
2MASS J00405912+4036536 &  10.246338 &  40.614906 & -10.819 &  -7.591 &  
1.57\\
2MASS J00410100+4037072 &  10.254204 &  40.618690 &  -9.944 &  -6.668 &  
1.67\\
2MASS J00410122+4037321 &  10.255102 &  40.625584 & -10.320 &  -7.055 &  
1.64\\
2MASS J00413945+4037416 &  10.414384 &  40.628242 &  -9.646 &  -6.392 &  
1.62\\
2MASS J00411799+4040170 &  10.324980 &  40.671394 & -10.198 &  -6.999 &  
1.52\\
2MASS J00414753+4040588 &  10.448079 &  40.683018 &  -9.645 &  -6.429 &  
1.55\\
2MASS J00405497+4044108 &  10.229067 &  40.736359 & -10.713 &  -7.413 &  
1.73\\
2MASS J00405619+4045461 &  10.234160 &  40.762817 & -11.136 &  -7.919 &  
1.55\\
2MASS J00421697+4047487 &  10.570731 &  40.796864 & -10.310 &  -6.976 &  
2.08\\
2MASS J00412184+4050090 &  10.341023 &  40.835857 & -10.957 &  -7.708 &  
1.61\\
2MASS J00414909+4050485 &  10.454579 &  40.846828 &  -9.736 &  -6.405 &  
1.86\\
2MASS J00402901+4050485 &  10.120903 &  40.846828 &  -9.907 &  -6.717 &  
1.50\\
2MASS J00424499+4051349 &  10.687467 &  40.859707 & -10.687 &  -7.432 &  
1.62\\
2MASS J00415135+4052139 &  10.463964 &  40.870552 &  -9.523 &  -6.315 &  
1.53\\
2MASS J00404426+4054039 &  10.184426 &  40.901096 &  -9.651 &  -6.452 &  
1.52\\
2MASS J00404649+4055220 &  10.193716 &  40.922802 & -10.666 &  -7.445 &  
1.56\\
2MASS J00424067+4059225 &  10.669477 &  40.989601 &  -9.542 &  -6.306 &  
1.58\\
2MASS J00424586+4059583 &  10.691123 &  40.999554 & -10.352 &  -7.136 &  
1.55\\
2MASS J00401171+4101138 &  10.048817 &  41.020523 & -10.045 &  -6.731 &  
1.78\\
2MASS J00404879+4102304 &  10.203307 &  41.041782 &  -9.769 &  -6.579 &  
1.50\\
2MASS J00440117+4104054 &  11.004878 &  41.068172 &  -9.585 &  -6.369 &  
1.55\\
2MASS J00413891+4104262 &  10.412128 &  41.073956 & -10.788 &  -7.459 &  
2.11\\
2MASS J00410868+4104383 &  10.286197 &  41.077316 & -10.902 &  -7.586 &  
2.17\\
2MASS J00414286+4107300 &  10.428614 &  41.125015 &  -9.872 &  -6.671 &  
1.52\\
2MASS J00434139+4109384 &  10.922468 &  41.160683 &  -9.630 &  -6.365 &  
1.64\\
2MASS J00412777+4109388 &  10.365709 &  41.160801 &  -9.681 &  -6.456 &  
1.56\\
2MASS J00412030+4111073 &  10.334590 &  41.185371 &  -9.927 &  -6.712 &  
1.55\\
2MASS J00434371+4111258 &  10.932143 &  41.190514 & -10.117 &  -6.831 &  
1.69\\
2MASS J00434135+4112137 &  10.922321 &  41.203815 & -10.265 &  -7.036 &  
1.57\\
2MASS J00433662+4112152 &  10.902587 &  41.204224 &  -9.793 &  -6.553 &  
1.59\\
2MASS J00405596+4112152 &  10.233187 &  41.204247 &  -9.646 &  -6.423 &  
1.56\\
2MASS J00434617+4112331 &  10.942393 &  41.209213 &  -9.952 &  -6.618 &  
1.88\\
2MASS J00434686+4112451 &  10.945257 &  41.212532 & -10.839 &  -7.615 &  
1.56\\
2MASS J00435338+4113016 &  10.972420 &  41.217136 & -11.022 &  -7.682 &  
1.95\\
2MASS J00435217+4114228 &  10.967414 &  41.239681 &  -9.583 &  -6.307 &  
1.67\\
2MASS J00424061+4115009 &  10.669212 &  41.250271 & -11.215 &  -7.952 &  
1.64\\
2MASS J00441547+4116071 &  11.064496 &  41.268646 &  -9.388 &  -6.179 &  
1.53\\
2MASS J00440575+4117199 &  11.023975 &  41.288864 & -10.343 &  -7.139 &  
2.43\\
2MASS J00440440+4117260 &  11.018359 &  41.290573 & -10.537 &  -7.197 &  
2.00\\
2MASS J00414297+4117300 &  10.429050 &  41.291691 & -10.946 &  -7.652 &  
1.71\\
2MASS J00414817+4118332 &  10.450715 &  41.309223 &  -9.873 &  -6.613 &  
1.63\\
2MASS J00441789+4119230 &  11.074565 &  41.323059 & -10.849 &  -7.612 &  
1.58\\
2MASS J00441402+4121420 &  11.058431 &  41.361691 & -10.121 &  -6.899 &  
1.56\\
2MASS J00442322+4121496 &  11.096773 &  41.363789 &  -9.974 &  -6.635 &  
1.93\\
2MASS J00442890+4121589 &  11.120435 &  41.366371 &  -9.809 &  -6.470 &  
1.93\\
2MASS J00443503+4123585 &  11.145990 &  41.399609 &  -9.559 &  -6.336 &  
1.56\\
2MASS J00415182+4124420 &  10.465925 &  41.411690 & -10.568 &  -7.342 &  
1.56\\
2MASS J00443680+4124465 &  11.153358 &  41.412933 & -11.078 &  -7.739 &  
2.02\\
2MASS J00443163+4125579 &  11.131823 &  41.432758 &  -9.911 &  -6.702 &  
1.53\\
2MASS J00405722+4127239 &  10.238433 &  41.456665 &  -9.710 &  -6.468 &  
1.59\\
2MASS J00443076+4128352 &  11.128191 &  41.476463 & -10.435 &  -7.217 &  
1.55\\
2MASS J00423161+4129146 &  10.631723 &  41.487389 & -10.476 &  -7.211 &  
1.64\\
2MASS J00445416+4129530 &  11.225678 &  41.498062 & -10.214 &  -6.989 &  
1.56\\
2MASS J00444232+4130556 &  11.176369 &  41.515465 &  -9.295 &  -6.074 &  
1.56\\
2MASS J00450163+4131148 &  11.256814 &  41.520779 & -11.079 &  -7.884 &  
1.51\\
2MASS J00445845+4131352 &  11.243563 &  41.526451 & -10.523 &  -7.288 &  
1.58\\
2MASS J00424172+4132501 &  10.673849 &  41.547260 & -10.530 &  -7.268 &  
2.32\\
2MASS J00424417+4133040 &  10.684061 &  41.551125 & -10.199 &  -6.861 &  
2.04\\
2MASS J00423133+4133103 &  10.630557 &  41.552876 & -10.480 &  -7.149 &  
2.10\\
2MASS J00451065+4133248 &  11.294379 &  41.556915 &  -9.826 &  -6.490 &  
1.90\\
2MASS J00441105+4133354 &  11.046080 &  41.559841 & -10.288 &  -7.059 &  
1.57\\
2MASS J00423774+4135395 &  10.657289 &  41.594326 &  -9.745 &  -6.453 &  
1.71\\
2MASS J00451348+4136308 &  11.306202 &  41.608582 & -10.546 &  -7.229 &  
1.79\\
2MASS J00425798+4137099 &  10.741592 &  41.619442 &  -9.524 &  -6.279 &  
1.60\\
2MASS J00430247+4137386 &  10.760300 &  41.627399 & -10.649 &  -7.389 &  
1.63\\
2MASS J00430142+4137423 &  10.755923 &  41.628422 & -10.393 &  -7.150 &  
1.60\\
2MASS J00430151+4137551 &  10.756292 &  41.631985 & -10.597 &  -7.297 &  
1.73\\
2MASS J00430776+4138124 &  10.782355 &  41.636780 & -10.488 &  -7.261 &  
1.57\\
2MASS J00451834+4139220 &  11.326427 &  41.656124 &  -9.708 &  -6.442 &  
1.64\\
2MASS J00432196+4141126 &  10.841507 &  41.686848 &  -9.919 &  -6.639 &  
1.68\\
2MASS J00445355+4141357 &  11.223136 &  41.693253 &  -9.497 &  -6.279 &  
1.55\\
2MASS J00454539+4142352 &  11.439165 &  41.709801 &  -9.465 &  -6.240 &  
1.56\\
2MASS J00451861+4143201 &  11.327579 &  41.722263 & -10.017 &  -6.784 &  
1.58\\
2MASS J00434203+4144032 &  10.925149 &  41.734238 &  -9.785 &  -6.493 &  
1.71\\
2MASS J00435943+4146429 &  10.997627 &  41.778587 &  -9.328 &  -6.131 &  
1.52\\
2MASS J00440219+4147493 &  11.009132 &  41.797043 & -10.116 &  -6.917 &  
1.52\\
2MASS J00441497+4148529 &  11.062387 &  41.814697 &  -9.513 &  -6.319 &  
1.51\\
2MASS J00425297+4149210 &  10.720725 &  41.822521 &  -9.685 &  -6.493 &  
1.51\\
2MASS J00440179+4149236 &  11.007496 &  41.823238 &  -9.561 &  -6.339 &  
1.56\\
2MASS J00440685+4149571 &  11.028579 &  41.832550 &  -9.483 &  -6.294 &  
1.50\\
2MASS J00445634+4149579 &  11.234758 &  41.832775 &  -9.519 &  -6.277 &  
1.59\\
2MASS J00454261+4150058 &  11.427546 &  41.834961 &  -9.937 &  -6.667 &  
1.65\\
2MASS J00442653+4150557 &  11.110566 &  41.848827 & -10.102 &  -6.771 &  
1.86\\
2MASS J00442081+4151597 &  11.086717 &  41.866611 & -11.269 &  -8.068 &  
1.52\\
2MASS J00443048+4152231 &  11.127036 &  41.873104 & -10.714 &  -7.451 &  
1.64\\
2MASS J00444915+4152273 &  11.204827 &  41.874271 & -11.053 &  -7.802 &  
1.61\\
2MASS J00443153+4153031 &  11.131396 &  41.884197 & -10.354 &  -7.142 &  
1.54\\
2MASS J00453263+4153356 &  11.385993 &  41.893234 &  -9.550 &  -6.362 &  
1.50\\
2MASS J00451817+4153503 &  11.325713 &  41.897324 &  -9.580 &  -6.387 &  
1.51\\
2MASS J00441502+4154102 &  11.062585 &  41.902859 &  -9.431 &  -6.186 &  
1.60\\
2MASS J00443323+4154382 &  11.138468 &  41.910625 &  -9.529 &  -6.255 &  
1.66\\
2MASS J00443446+4154487 &  11.143602 &  41.913532 &  -9.557 &  -6.246 &  
1.77\\
2MASS J00444960+4154503 &  11.206699 &  41.913990 &  -9.482 &  -6.281 &  
1.52\\
2MASS J00452458+4155249 &  11.352445 &  41.923588 &  -9.749 &  -6.522 &  
1.57\\
2MASS J00443676+4155300 &  11.153187 &  41.925011 &  -9.749 &  -6.515 &  
1.58\\
2MASS J00453432+4158096 &  11.393025 &  41.969357 &  -9.835 &  -6.617 &  
1.55\\
\end{supertabular}
\end{center}

\vspace{1cm}
\begin{center}
\topcaption{Very luminous AGB stars (carbon star candidates) in M33}
\label{m33}
\tablefirsthead{
  \hline
  \hline
  2MASS ID & RA   & Dec   & $M_{K}$ & $M_{\rm bol}$ & $J-K_s$ \\
           &(deg) & (deg) & (mag)   & (mag) & (mag)\\
  \hline}
  \tablehead{
  \hspace{-1cm}table continuation\\
    \hline
  2MASS ID & RA   & Dec   & $M_{K}$ & $M_{\rm bol}$ & $J-K_s$ \\
           &(deg) & (deg) & (mag)   & (mag) & (mag)\\
  \hline}
  \tabletail{\hline}
\begin{supertabular}{lrrrrr}
2MASS J01331769+3021253 &  23.323723 &  30.357054 &  -9.970 &  -6.735 &  
1.58\\
2MASS J01335587+3027285 &  23.482809 &  30.457930 & -10.630 &  -7.341 &  
1.70\\
2MASS J01330404+3013225 &  23.266844 &  30.222937 &  -9.714 &  -6.399 &  
1.78\\
2MASS J01323279+3030250 &  23.136661 &  30.506952 &  -9.829 &  -6.489 &  
1.95\\
2MASS J01340022+3040475 &  23.500920 &  30.679874 & -10.369 &  -7.064 &  
1.75\\
2MASS J01335459+3040262 &  23.477486 &  30.673952 & -10.671 &  -7.340 &  
1.86\\
2MASS J01340697+3047539 &  23.529077 &  30.798306 & -10.084 &  -6.803 &  
1.68\\
2MASS J01340090+3034465 &  23.503776 &  30.579611 &  -9.948 &  -6.759 &  
1.50\\
2MASS J01333366+3033204 &  23.390251 &  30.555672 &  -9.750 &  -6.472 &  
1.67\\
2MASS J01341556+3052499 &  23.564865 &  30.880529 & -10.932 &  -7.592 &  
2.01\\
2MASS J01333418+3041271 &  23.392448 &  30.690870 & -10.598 &  -7.281 &  
1.79\\
2MASS J01335855+3033525 &  23.493967 &  30.564590 &  -9.540 &  -6.331 &  
1.54\\
2MASS J01335488+3038537 &  23.478687 &  30.648275 &  -9.900 &  -6.594 &  
1.75\\
2MASS J01333862+3032254 &  23.410958 &  30.540401 & -10.196 &  -6.891 &  
1.75\\
2MASS J01334194+3038565 &  23.424778 &  30.649031 & -10.520 &  -7.301 &  
1.55\\
2MASS J01334389+3032246 &  23.432910 &  30.540171 & -10.020 &  -6.763 &  
1.63\\
2MASS J01335316+3038389 &  23.471534 &  30.644152 & -10.157 &  -6.897 &  
1.63\\
2MASS J01333357+3031599 &  23.389879 &  30.533327 & -10.884 &  -7.555 &  
2.11\\
2MASS J01334172+3050390 &  23.423867 &  30.844172 &  -9.993 &  -6.656 &  
1.91\\
2MASS J01332901+3042541 &  23.370907 &  30.715054 &  -9.587 &  -6.324 &  
1.64\\
2MASS J01340652+3054510 &  23.527172 &  30.914177 & -10.135 &  -6.795 &  
2.00\\
2MASS J01334812+3039301 &  23.450541 &  30.658377 &  -9.842 &  -6.617 &  
1.56\\
2MASS J01340055+3059387 &  23.502324 &  30.994099 &  -9.748 &  -6.462 &  
1.69\\
2MASS J01341459+3033543 &  23.560816 &  30.565096 & -10.211 &  -6.985 &  
1.57\\
2MASS J01355238+3051030 &  23.968257 &  30.850855 &  -9.520 &  -6.295 &  
1.56\\
2MASS J01350167+3054534 &  23.756997 &  30.914835 & -10.324 &  -7.042 &  
1.68\\
2MASS J01340739+3035173 &  23.530827 &  30.588150 &  -9.636 &  -6.343 &  
1.71\\
2MASS J01323260+3030567 &  23.135842 &  30.515766 &  -9.592 &  -6.394 &  
1.52\\
2MASS J01323983+3038170 &  23.165985 &  30.638075 &  -9.556 &  -6.308 &  
1.61\\
2MASS J01330147+3033361 &  23.256127 &  30.560051 & -10.030 &  -6.788 &  
1.60\\
2MASS J01321356+3040033 &  23.056537 &  30.667587 &  -9.520 &  -6.327 &  
1.51\\
\end{supertabular}
\end{center}


\begin{thebibliography}{}

%\bibitem[1964]{arp}
%  Arp, H., 1964, ApJ, 139, 1045

\bibitem[2005]{battinelli1}
  Battinelli, P., Demers, S., 2005, A\&A, 430, 905

\bibitem[2003]{battinelli2}
  Battinelli, P., Demers, S., Letarte, B., 2003, AJ, 125, 1298

\bibitem[1982]{beck}
  Beck, R., Gr\"{a}ve, R., 1982, A\&A, 105, 192

\bibitem[1988]{bessel}
  Bessell, M. S., Brett, J. M., 1988, PASP, 100, 1134

\bibitem[1984]{bessel2}
  Bessell, M. S., Wood, P. R., 1984, PASP, 96, 247

\bibitem[2004]{block}
  Block, D. L., Freeman, K. C., Jarrett, T. H., Puerari, I., Worthey, G.,
Combes, F., Groess, R., 2004, A\&A, 425, L37

\bibitem[2000]{bonnarel}
  Bonnarel, F., Fernique, P., Bienaym\'e, O., Egret, D., Genova, F., Louys, 
M.,
Ochsenbein, F., Wenger, M., Bartlett, J. G., 2000, A\&AS, 143, 33

%\bibitem[1991]{braun}
%  Braun, R., 1991, ApJ, 372, 54

\bibitem[1996]{brewer1}
  Brewer, J. P., Richer, H. B., Crabtree, D. R., 1996, AJ, 112, 491

\bibitem[1995]{brewer2}
  Brewer, J. P., Richer, H. B., Crabtree, D. R., 1995, AJ, 109, 2480

\bibitem[1984]{brinks}
  Brinks, E., Shane, W. W., 1984, A\&AS, 55, 179

\bibitem[2004]{cioni}
  Cioni, M. R. L., Habing, H. J., Loup, C., Epchtein, N., Deul, E., 2004, 
Msngr, 115, 22

\bibitem[2005]{davidge1}
  Davidge, T. J., Olsen, K. A. G., Blum, R., Stephens, A. W., Rigaut, F. 
2005,
AJ, 129, 201

\bibitem[2003]{davidge2}
  Davidge, T. J., 2003, ApJ, 597, 289

\bibitem[2003]{demers}
  Demers, S., Battinelli, P., Letarte, B., 2003, A\&A, 410, 795

\bibitem[1994]{devereux}
  Devereux, N. A., Price, R., Wells, L. A., Duric, N., 1994, AJ, 108, 1667

\bibitem[2001]{egan}
  Egan, M. P., Van Dyk, S. D., Price, S. D., 2001, AJ, 122, 1844
  
\bibitem[1998]{frost}
  Frost, C. A., Cannon, R., C., Lattanzio, J. C., Wood, P. R., Forestini, 
M.,
1998, A\&A, 332, L17

\bibitem[2002]{groenewegen}
  Groenewegen, M. A. T., 2002, astro-ph/0208449

\bibitem[1998]{haas}
  Haas, M., Lemke, D., Stickel, M., Hippelein, H., Kunkel, M., 
Herbstmeier, U.,
Mattila, K., A\&A, 1998, 338, L33

\bibitem[1992]{hodge}
  Hodge, P., 1992, The Andromeda Galaxy, Kluwer Academic Publishers, 
Dordrecht

\bibitem[1990]{hughes}
  Hughes, S. M. G., Wood, P. R., 1990, AJ, 99, 784

\bibitem[2004]{kale}
  Kale, S., Vijayaraman, T.~M., Kembhavi, A., Krishnan, P.~R., Navelkar, 
A.,
  Hegde, H., Kulkarni, P., \& Balaji, K.~D.\ 2004, Astronomical Society of
  the Pacific Conference Series, 314, 350 

\bibitem[2001]{kontizas}
  Kontizas, E., Dapergolas, A., Morgan, D. H., Kontizas, M., 2001, A\&A, 
369,
932

\bibitem[1993]{kontizas2}
  Kontizas, M., Kontizas, E., Michalitsianos, A. G., 1993, A\&A, 269, 107

\bibitem[2002]{lee}
  Lee, M. G., Kim, M., Sarajedini, A., Geisler, D., Gieren, W., 2002, ApJ, 
565,
959

\bibitem[2003]{marigo}
  Marigo, P., Girardi, L., Chiosi, C., 2003, A\&A, 403, 225

\bibitem[1995]{massey}
  Massey, P., Lang, C. C., Degioia-Eastwood, K., Garmany, C. D., 1995, 
ApJ, 438, 188

\bibitem[1998]{montegriffo}
  Montegriffo, P., Ferraro, F. R., Origlia, L., Fusi Pecci, F., 1998, 
MNRAS, 297, 872

\bibitem[1995]{morgan}
  Morgan, D. H., Hatzidimitriou, D., 1995, A\&AS, 113, 539

\bibitem[2000]{nikolaev}
  Nikolaev S. and Weinberg M. D., 2000, ApJ, 542, 804

\bibitem[2000]{ochsenbein}
  Ochsenbein, F., Bauer, P., Marcout, J., 2000, A\&AS, 143, 23

\bibitem[2004]{padovani}
  Padovani, P., AVO, 2004, AAS, 20513006

\bibitem[1978]{pellet}
  Pellet, A., Astier, N., Viale, A., Courtes, G., Maucherat, A., Monnet, 
G.,
Simien, F., 1978, A\&AS, 31, 439

\bibitem[1993]{rebeirot}
  Rebeirot, E., Azzopardi, M., Westerlund, B. E., 1993, A\&AS, 97, 603

\bibitem[1989]{richer}
  Richer, H. B., 1989, IAU Coll. 106, (Cambridge University Press), 35

\bibitem[2005]{rowe}
  Rowe, J. F., Richer, H. B., Brewer, J. P., Crabtree, D. R., 2005, AJ, 
129, 729

\bibitem[1981]{sofue}
  Sofue, Y., Kato, T., 1981, Publ. Astron. Soc. Japan, 33, 449

\bibitem[2005]{thilker}
  Thilker, D. A., Hoopes, C. G., Bianchi, L., Boissier, S., Rich, R. M., 
Seibert,
M., Friedman, P. G., Rey, S. C., Buat, V.,
     Barlow, T. A. and 18 coauthors  2005, ApJ, 619L, 67

\bibitem[1991]{van}
  van den Bergh, S., 1991, PASP, 103, 1053

\bibitem[1991]{vanLoon}
  van Loon, J. Th., Zijlstra, A. A., Kaper, L., Gilmore, G. F., Loup, C.,
Blommaert, J. A. D. L., 2001, A\&A, 368, 239

\bibitem[1992]{wainscoat}
  Wainscoat, R. J., Cohen, M., Volk, K., Walker, H. J., Schwartz, D. E.,
     1992, ApJS, 83, 111

\bibitem[1988]{walterbos}
  Walterbos, R. A. M., Kennicutt, R. C, Jr., 1988, A\&A, 198, 61

\bibitem[1991]{wilson}
  Wilson, C. D., 1991, AJ, 101, 1663

\bibitem[1985]{wood}
  Wood, P. R., Bessell, M. S., Paltoglou, G., 1985, ApJ, 290, 477

\end{thebibliography}
\end{document}